\documentclass[default,iicol]{sn-jnl}

\usepackage{graphicx}%
\usepackage{multirow}%
\usepackage{amsmath,amssymb,amsfonts}%
\usepackage{amsthm}%
\usepackage{mathrsfs}%
\usepackage[title]{appendix}%
\usepackage{xcolor}%
\usepackage{textcomp}%
\usepackage{manyfoot}%
\usepackage{booktabs}%
\usepackage{algorithm}%
\usepackage{algorithmicx}%
\usepackage{algpseudocode}%
\usepackage{listings}%
\usepackage[normalem]{ulem}

\newcommand{\f}{{ f}}
\newcommand{\g}{{ g}}
\newcommand{\Fo}{{ {\mathcal F}_1}}
\newcommand{\Ft}{{{\mathcal F}_2}}

\begin{document}
\newcommand{\matrixel}[3]{\left< #1 \vphantom{#2#3} \right| #2 \left| #3 \vphantom{#1#2} \right>}
\newcommand{\abs}[1]{\left| #1 \right|} 
\newcommand{\avg}[1]{\left< #1 \right>}

\title[$B$-physics from Lattice Gauge Theory]{$B$-physics from Lattice Gauge Theory}

\author*[1]{\fnm{J. Tobias} \sur{Tsang}}\email{j.t.tsang@cern.ch}
\author[2]{\fnm{Michele} \sur{Della Morte}}

\affil[1]{\orgdiv{Department of Theoretical Physics}, \orgname{CERN}, \orgaddress{\street{Esplanade des Particules 1}, \city{Geneva}, \postcode{1211}, \country{Switzerland}}}
\affil[2]{IMADA Department, University of Southern Denmark, Campusvej 55, Odense, 5230, Denmark}

\abstract{We discuss the main issues in dealing with heavy quarks on the lattice
  and shortly present the different approaches used.  We discuss a selection of
  computations covering first the $b$-quark mass and the $B_{(s)}$ meson decay
  constants as the consolidated results (neglecting isospin breaking
  corrections).  In the second part we consider recent calculations of form
  factors for tree-level semileptonic decays with emphasis on the tensions
  between the results produced by different collaborations. We propose benchmark
  quantities and tests suited to investigate the origin of such
  tensions. Finally, we review computations of the bag parameters parameterising
  neutral meson mixing and provide an overview on a few recent developments in
  the field.
}


\maketitle

\section{Introduction}\label{sec1}
The large $b$-quark mass and the comparably long lifetime of $B_{(s)}$ mesons,
allow for a plethora of experimental observables which can be measured precisely
(see for example Refs.~\cite{Cerri:2018ypt} and~\cite{Belle-II:2018jsg} for a
discussion of the opportunities at LHC and Belle II, respectively).  If
precision Standard Model (SM) predictions are available, these can be used to
test the SM as well as to search for and constrain New Physics (NP) beyond the
Standard Model (BSM). In addition to direct comparisons between experiment and
theory, the self-consistency of the SM can be tested by over-constraining the
CKM matrix~\cite{Cabibbo:1963yz,Kobayashi:1973fv} and testing its unitarity.

Lattice QCD is the only known tool to provide non-perturbative, \emph{ab
  initio}, systematically improvable precision predictions. In lattice QCD a
finite Euclidean space-time volume ($L$) is discretised (with lattice spacing
$a$) and a representative \emph{ensemble} of the gauge field configurations is
sampled via Monte Carlo methods. On these configurations, correlation functions
are computed from which hadronic masses and matrix elements can be
extracted. This is repeated for multiple choices of the simulation parameters
($a$, $L$ and the quark masses $am_q$) and to make contact with experiment the
observables are inter/extrapolated to the physical world ($a\to 0, L\to \infty,
am_q \to am_q^\mathrm{phys}$). It is noteworthy, that since lattice QCD
simulations take bare quark masses as inputs, predictions at unphysical quark
masses can be made, which can be used to test effective field theories (EFTs)
such as Chiral Perturbation Theory ($\chi$PT) or Heavy Quark Effective Theory
(HQET).  Modern simulations include contributions from sea effects of two
degenerate light quarks and the strange quark ($N_f = 2+1$) and the charm quark
($N_f = 2+1+1$).

In recent years lattice QCD calculations of hadronic observables containing a
$b$ quark have made huge advances with precision predictions with all systematic
uncertainties under control for several quantities.\footnote{For a recent lattice review
we refer to the latest plenary proceedings at a Lattice
Conference~\cite{Kaneko:2023kxx}.} Here, we comment on some particular
challenges when simulating heavy quarks (Sec.~\ref{sec:challenges}) before
summarising the status of computations of the $b$-quark mass and leptonic decay
constants (Sec.~\ref{sec:mBfB}), semileptonic decay form factors
(Sec.~\ref{sec:semi}) and neutral meson mixing (Sec.~\ref{sec:mix}). In
Sec.~\ref{sec:recent} we comment on recent, more exploratory developments before
concluding in Sec~\ref{sec:conc}.

\section{Lattice challenges for heavy quarks \label{sec:challenges}}
Including heavy quarks, such as the $b$ quark, in lattice simulations of QCD
poses a multi-scale problem and currently all approaches still rely on the use
(even though possibly at different stages) of EFTs, typically
HQET~\cite{Eichten:1987xu,Eichten:1989zv} or Non-Relativistic QCD
(NRQCD)~\cite{Caswell:1985ui}.  The infrared scale is set by the lattice extent
$L$ and the dynamics of the light degrees of freedom, such as the pions, should
not be distorted by the finite size of the lattice. The standard requirement is
to have $m_\pi L > 4$, with $m_\pi$ the pion mass.  The ultraviolet scale is
instead set by the lattice spacing $a$, the finest resolution in the system. In
order to properly resolve the propagation of the heavy degrees of freedom, such
as the $b$ quark, the mass $m_b$ should be far from the cutoff $1/a$ or in other
words $am_b $ should be smaller than one. Substituting the physical value for
$m_b$, one concludes that lattices with $L/a$ significantly larger than 100 are
needed to keep both finite size and discretisation (or cutoff) effects under
control.  While such simulations are becoming feasible with current machines and
algorithms, in order to perform a controlled continuum extrapolation ($a \to 0$)
at the $b$-quark mass one will need simulations at even finer resolutions. This
might require new algorithms in order to ensure an ergodic sampling of the
configuration space (see Ref.~\cite{Luscher:2011kk}).

In this context EFTs are used essentially in two, non-exclusive, ways.  Either
they are implemented directly on the lattice by (formally) expanding physical
quantities in inverse powers of the scale associated to the heavy degrees of
freedom (e.g., the $b$-quark mass), or they are used to extrapolate results
obtained for heavier-than-charm-quark masses to the $b$-quark mass. In a
hard-cutoff regularisation, such as the lattice regularisation, the first
approach often produces power-divergences (divergences in inverse powers of the
lattice spacing), due to the mixing between operators of different
dimensions. Those indeed naturally appear in an expansion in terms of couplings
of negative dimension such as $1/m_b$.  As pointed out in
Refs.~\cite{Sommer:2006sj, DellaMorte:2007ny}, these power divergences need to
be removed non-perturbatively if one wants to perform a continuum limit
extrapolation.

In practice one considers two EFTs, one being HQET (or NRQCD) and the other the
Symanzik Effective Theory (SyEFT)~\cite{Symanzik:1983dc,Symanzik:1983gh},
therefore implementing the O($a$) improvement programme to systematically reduce
cutoff effects.  In SyEFT higher-dimensional operators are introduced depending
on the symmetries of the lattice action. Dimensions are compensated by powers of
the lattice spacing and the coefficients are tuned in order to remove the
leading lattice artifacts.  Roughly speaking, different approaches implement the
two EFTs in different order.

In O($a$)-improved HQET, one considers the static Lagrangian density
\begin{equation}
  \,\,{\cal{L}}_{stat}(x)\!=\!\overline{\psi}_h(x) D_0 \psi_h(x) \,; \,\,\,\, \frac{1+\gamma_0}{2} \psi_h\! = \!\psi_h
    \end{equation}
for the heavy quarks. This locally preserves the heavy flavour number and is
symmetric under continuous SU(2) rotations in Dirac space
\begin{equation}
    \psi_h \to V(\vec{\phi}) \psi_h\;, \quad {\rm with} \quad V(\vec{\phi})=e^{\phi_i \epsilon_{ijk} \sigma_{jk}}\;,
\end{equation}
where $\vec{\phi}$ is the transformation parameter. Such symmetries are only
realised in the infinite mass limit; they are not symmetries of QCD and are
broken by $1/m_b$ corrections.  At the static order a term $\delta
m\,\overline{\psi}(x)\psi(x)$ can be added to the Lagrangian with $\delta m$
being a power divergent (as $1/a$) parameter producing a shift in the energy
levels of heavy-light systems. This is the first occurrence of the power
divergences mentioned above.  The subtraction of this divergence can be
performed through a non-perturbative matching between HQET and QCD, as discussed
in Ref.~\cite{Heitger:2003nj} for the example of the computation of the
$b$-quark mass.

Higher orders in $1/m_b$ appear with operators of dimension 5 and
larger. Including those in the Lagrangian would produce a non-renormalisable
theory and therefore such corrections are treated order-by-order as space-time
volume insertions in correlation functions computed in the static theory (see
Ref.~\cite{DellaMorte:2006chd} for a general discussion, applied to the
calculation of the $b$-quark mass including $1/m_b$ corrections). Similarly,
when considering matrix elements, local operators have a static expression and
higher order (and higher dimensional) corrections that appear with appropriate
powers of $1/m_b$.  Those have to be treated together with the higher order
terms in the Lagrangian and a framework for a complete and non-perturbative
matching of the action and the vector and axial heavy-light currents between QCD
and HQET has been put forward in Ref.~\cite{DellaMorte:2013ega}.

The Symanzik improvement programme is implemented in a straightforward way for
the action and operators. As only the static action is directly simulated
(higher orders being treated as insertions) the symmetries of the static theory
can be used in classifying the mixing between operators of different dimensions
for the O($a$)-improvement.

Concerning the quantities reviewed here, results from non-perturbative HQET are
available for the $B_s \to K \ell \nu$ form factors~\cite{Bahr:2016ayy,
  Bahr:2019eom}, the $b$-quark mass~\cite{DellaMorte:2006chd,
  Bernardoni:2013xba}, $B_{(s)}$-meson decay constants~\cite{DellaMorte:2007ny,
  Blossier:2010mk, ALPHA:2014lwy} and mixing parameters~\cite{DellaMorte:2004wn,
  RBC:2007bua}. Since most recent results use different formulations for the $b$
quark, we will not further discuss this approach here.

By interchanging the order in which the Symanzik expansion and the Heavy Quark
expansion are performed, one is led to the relativistic heavy-quark formulation
such as the one introduced in Ref.~\cite{El-Khadra:1996wdx}, which is now known
as the \emph{Fermilab action}. The main property is that the coefficients of the
higher dimensional operators appearing through the Symanzik improvement
programme are allowed to depend explicitly on the heavy-quark mass $m_h$.  In
this way the relativistic heavy-quark actions, at fixed lattice spacing,
interpolate between the massless limit and the infinite-mass (static) limit.  In
particular, this implies that for $am_h \gg 1$ one expects to recover the
power-like divergences of lattice HQET. The starting point is the anisotropic
clover action density
\begin{eqnarray}
    &&{\cal{L}}_{Fermilab}(x)=a^4\overline{\psi}(x) \left( m_0 + \gamma_0 D_0 + \zeta \vec{\gamma}\cdot\vec{D} \right. \nonumber \\
    &&\left. -\frac{a}{2}D_0^2 - \frac{a}{2}\zeta\vec{D}^2 +\frac{ia}{4}c_{SW}\sigma_{\mu\nu} F_{\mu\nu}
    \right)\psi(x)  \;,
\label{Fermact}
\end{eqnarray}
where the anisotropy parameter $\zeta$, the clover coefficient $c_{SW}$ and the
mass parameter $m_0$ are tuned to reproduce experimental values for $B$-mesons
spectral quantities such as the dispersion relation and the vector-pseudoscalar
splitting~\cite{El-Khadra:1996wdx,Aoki:2001ra,Christ:2006us}.  The heavy-quark
symmetries emerge naturally in the action in eq.~\eqref{Fermact} and therefore
HQET can be used to model and estimate cutoff
effects~\cite{Kronfeld:2000ck,Harada:2001fi,Harada:2001fj}. A variant of the
Fermilab action, where the parameters $m_0a$, $\zeta$ and $c_{P}$ (a generalised
version of $c_{SW}$) are tuned non-perturbatively~\cite{RBC:2012pds}, is known
as the \emph{Relativistic Heavy Quark} (RHQ)~\cite{Christ:2006us,Lin:2006ur}
action.

Other effective approaches have been used to treat heavy quarks on the
lattice. NRQCD is formally an expansion of QCD in powers of the heavy-quark
velocity $v$ and a lattice version has been introduced in
Refs.~\cite{Thacker:1990bm,Lepage:1992tx}. Leading operators as well as
operators suppressed by O$(v^2)$ are included in typical applications together
with operators needed for O$(a^2)$-improvement. The expression for the lattice
action is rather lengthy and can be found in
Ref.~\cite{Lepage:1992tx}. Operators up to dimension 7 appear at the order
mentioned above. Dimensions are compensated by powers of the heavy-quark mass
and the coefficients are determined through matching with QCD, typically
performed at tree-level or at most at one-loop.\footnote{Recently, a calculation
  with a non-perturbative tuning of the NRQCD action parameters has become
  available~\cite{Hudspith:2023loy}, however this is not the standard in the
  existing literature.} NRQCD is hence clearly non-renormalisable by power
counting and the continuum limit at fixed heavy-quark mass does not exist. The
accuracy that can be reached depends on the existence of a window in the lattice
spacing where both cutoff effects (positive powers of $a$) and power-like
divergences (negative powers of $a$) are under control. In actual computations
the precision is at the percent level and the approach is becoming less and less
used, since it would be computationally quite demanding to improve it (by
including operators of even higher dimension and performing the matching at high
loop orders).

Finally, in the most recent applications, heavy quarks are discretised using
regularisations originally introduced for the light flavours. This is often
referred to as \emph{fully relativistic formulations} and is possible because of
the very fine lattice spacings that can be reached in modern simulations (around
0.04 fm) and because of the use of highly improved actions, where mass-dependent
cutoff effects start at O($(am_h)^2$). Since the $b$-quark mass typically only
satisfies $am_b \lesssim 1$ for the finest lattice spacing, HQET is still used
for the simultaneous continuum and heavy-mass extrapolations.

The HPQCD Collaboration has introduced the use of highly improved staggered
quarks (HISQ action)~\cite{Follana:2006rc} at very fine lattice spacings and for
$b$ quarks in~\cite{McNeile:2011ng}. The first study concerned the $B_s$-meson
decay constant but by now the method has been applied to a variety of quantities
that will be discussed in the following.

In a less direct approach, the ETM Collaboration has been using automatically
O($a$)-improved twisted mass fermions with masses in the heavier-than-charm
region, together with results in the static approximation to interpolate to the
$b$-quark mass. The main implementation is through the ratio
method~\cite{ETM:2009sed}, where suitable ratios of heavy-light quantities, for
different heavy-quark mass, are constructed such that they possess a well
defined static limit (typically equal to 1).  $B$-physics quantities are then
obtained as an interpolation (rather than an extrapolation) between the static
value and the results of the simulations around the charm.

Results from the use of relativistic heavy-quark actions or fully relativistic
actions represent the majority of recent results, including those that we are
going to review here in the remaining part of this contribution.

\section{The $b$-quark mass and leptonic decay constants\label{sec:mBfB}}
\subsection{$b$-quark mass}
\begin{figure}
  \includegraphics[width=\columnwidth]{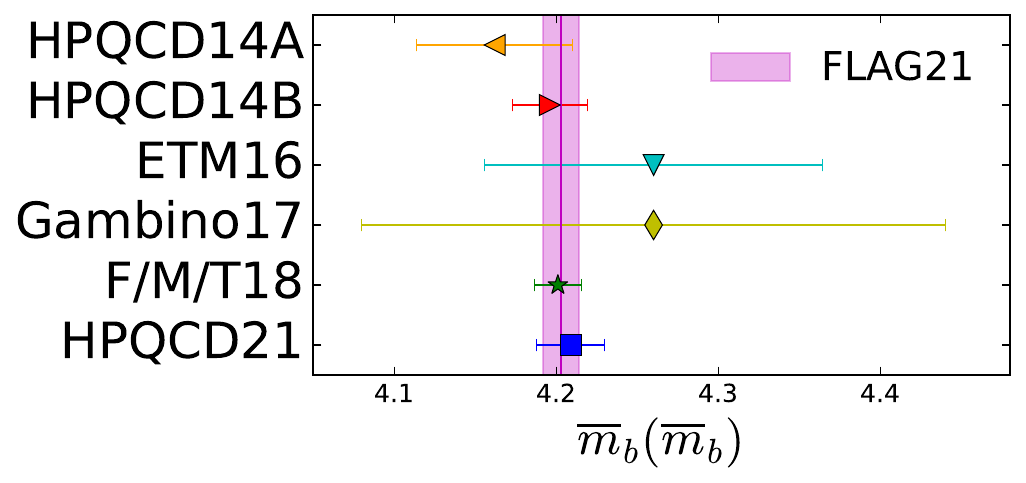}
  \caption{Summary of recent results for the $b$-quark mass in the
    $\overline{\text{MS}}$ scheme for calculations including the dynamical
    effects of $2+1+1$ flavours. Individual results shown are
    HPQCD14A~\cite{Chakraborty:2014aca}, HPQCD14B~\cite{Colquhoun:2014ica},
    ETM16~\cite{ETM:2016nbo}, Gambino17~\cite{Gambino:2017vkx},
    F/M/T18~\cite{FermilabLattice:2018est}, and
    HPQCD21~\cite{Hatton:2021syc}. The magenta band corresponds to the
    recommended value from the FLAG
    collaboration~\cite{FlavourLatticeAveragingGroupFLAG:2021npn} based on these
    results.}
  \label{fig:mb}
\end{figure}

The $b$-quark mass is a parameter of QCD and hence, from the theoretical point
of view, knowing its value is of fundamental importance. Lattice determinations
of the $b$-quark mass are three times more precise than continuum
ones~\cite{ParticleDataGroup:2022pth} and they drive the accuracy on the
parameter. In the latest edition of the FLAG
review~\cite{FlavourLatticeAveragingGroupFLAG:2021npn} the values
\begin{equation}
\overline{m}_b(\overline{m}_b)=4.203(11) \, {\rm GeV}\;,
\end{equation}
for the mass in the $\overline{\rm MS}$ scheme (illustrated as the magenta band
in Fig.~\ref{fig:mb}), and
\begin{equation}
M_b^{\rm RGI}=6.934(58)\, {\rm GeV}\;,
\end{equation}
for the Renormalisation Group Invariant (RGI) mass, are quoted using results
with $2+1+1$ dynamical flavours.  As a remark, in the RGI case the error (which
is around one percent) is dominated by the RG-evolution required in the
definition and therefore by the uncertainty on the $\Lambda$-parameter of QCD.

The most recent computations of the $b$-quark mass entering the FLAG average for
$N_f=2+1+1$ are shown in Fig.~\ref{fig:mb}. These results use the ratio method
described above~\cite{ETM:2016nbo,Gambino:2017vkx} or tune the heavy-quark mass,
possibly in some conveniently defined intermediate scheme, in order to reproduce
(or extrapolate to) the $B_{(s)}$-meson mass~\cite{FermilabLattice:2018est} or
extract the $b$-quark mass from the analysis of moments of heavy current-current
correlation
functions~\cite{Chakraborty:2014aca,Colquhoun:2014ica,Hatton:2021syc}. The
latter method is rather novel and has already produced some of the most precise
results and hence deserves a more detailed discussion.  The idea is first
introduced in Ref.~\cite{Chakraborty:2014aca} and consists of computing moments
\begin{equation}
    G_n=\sum_t (t/a)^n G(t) \;,
\end{equation}
of the zero-momentum two-point function $G(t)$ of the heavy pseudoscalar density
$j_5=am_{0h} \overline{\psi}_h \gamma_5 \psi_h$, where $am_{0h}$ is the bare
heavy-quark mass. More precisely, in order to reduce discretisation errors, one
computes the ratios $\tilde{R}_n$ defined as
\begin{equation}
  \tilde{R}_n = 
  \begin{cases}
    G_4/G_4^{(0)} &\quad {\rm for}\; n=4 \;,\\
    \frac{1}{m_{0h}} \left(G_n/G_n^{(0)} \right)^{1/(n-4)} &\quad {\rm for} \; n\geq 6 \;,
    \end{cases}
\end{equation}
with $G_n^{(0)}$ being the lowest perturbative order in the expansion of the
correlation function.  The key observation is that for low values of $n$ the
moments are short-distance dominated and they become more and more perturbative
as the quark mass is increased. By matching the lattice results to the continuum
perturbative prediction, one can simultaneously extract the heavy-quark mass in
the $\overline{\rm MS}$ scheme and the strong coupling $\alpha_s$. The
coefficients in the expansion of $\tilde{R}_n$, for low $n$, are known to order
$\alpha_s^3$
included~\cite{Chetyrkin:2006xg,Boughezal:2006px,Maier:2008he,Maier:2009fz,Kiyo:2009gb},
and the method can therefore provide rather accurate estimates of the strong
coupling and the $b$-quark mass.

\begin{figure}
  \hspace{-.5cm}
  \includegraphics[width=.25\textwidth]{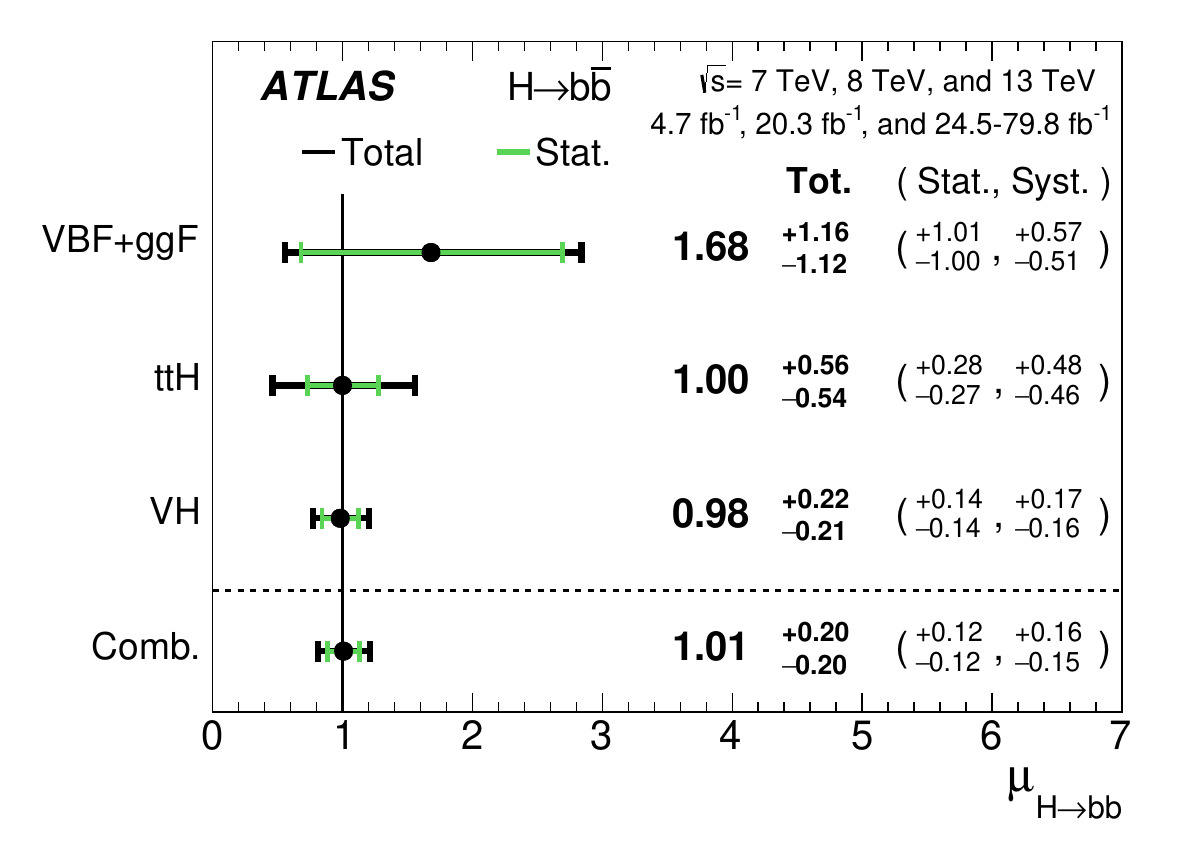}
  \includegraphics[width=.235\textwidth]{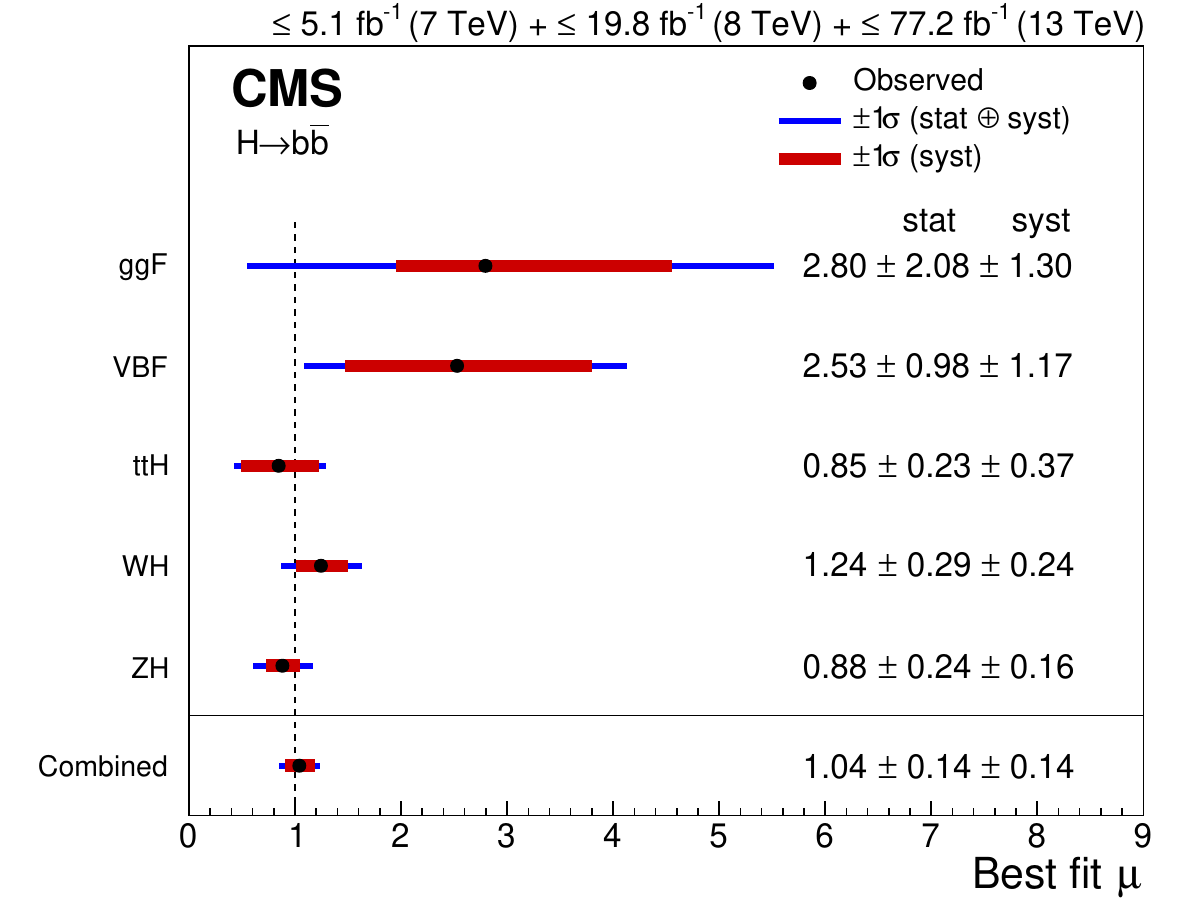}\\
  \caption{Signal strengths for $H\to b\overline{b}$ from ATLAS (left) and CMS (right). Figure from Refs.~\cite{CMS:2018nsn,ATLAS:2018kot}.}
  \label{fig:bbstrength}
  
\end{figure}

From the phenomenological point of view a precise knowledge of the $b$-quark
mass is required for testing the properties of the SM-like Higgs particle.  Its
dominant decay mode is $b\overline{b}$, with a branching ratio of about
60\%. The theoretical estimate of that depends quadratically on the $b$-quark
mass, which therefore contributes about 2\% to the total uncertainty. The signal
strengths (ratio of experimental to theoretical estimates) based on CMS and
ATLAS data~\cite{CMS:2018nsn,ATLAS:2018kot,ATLAS:2020fcp} are consistent with
the SM with an error of about 10\%. The PDG~\cite{ParticleDataGroup:2022pth}
averages the results to a signal strength of $0.99(12)$, the error being
currently dominated by the experimental uncertainties. Summary plots for the
different production channels, taken from Refs.~\cite{CMS:2018nsn,ATLAS:2018kot}
are shown in Fig.~\ref{fig:bbstrength}.

\subsection{Decay constants}
Within the Weak Effective Hamiltonian framework, and neglecting electromagnetic interactions, the  
decay rate for tree-level process $B^+\to \ell^+ \nu_\ell$ can be expressed as
\begin{equation}
\,\hspace{-.1cm}\Gamma(B\to \ell \nu)= \frac{m_B}{8\pi} G_F^2 f_B^2 |V_{ub}|^2m_\ell^2\!\left(1-\frac{m_\ell^2}{m_B^2}\right).\hspace{-.25cm}
\end{equation}
Similarly, the decay rate for the loop-mediated process $B_q^0\to \ell^+ \ell^-$
($q=d,s$) depends on the product $|V_{tb}^*V_{tq}|^2$ and the decay constant
$f_{B_q}^2$.  These hadronic parameters are defined by the QCD matrix element ($q=u,d,s,c$)
\begin{equation}
    \langle 0 | A_{bq}^\mu| B_q(p) \rangle = i f_{B_q} p^\mu  \;\; {\rm and} \; A^\mu_{bq}=\overline{b}\gamma_\mu\gamma_5 q\;,
\end{equation}
where the left-hand-side of the first equation is exactly what is computed on
the lattice. By combining experimental measurements with results for the decay
constants one can therefore obtain exclusive determinations of CKM matrix
elements such as $\abs{V_{ub}}$.  The measured channel is $B^-\to \tau^-
\overline{\nu}$ with uncertainties around 20\% on the branching ratio from the
two experiments Belle and BaBar~\cite{Belle:2015odw,BaBar:2012nus} and these
results show a tension a bit below the level of two combined standard
deviations. The averages of lattice results for the decay constants have an
uncertainty below one percent (for
$N_f=2+1+1$)~\cite{FlavourLatticeAveragingGroupFLAG:2021npn}, hence the
determinations of CKM matrix elements from leptonic decays are currently
dominated by the experimental error.  In that sense $B$-mesons decay constants
have become a benchmark computation for new methods dealing with heavy
quarks on the lattice.  A much more precise exclusive estimate of $\abs{V_{ub}}$
is extracted from the semileptonic channel $B \to \pi \ell \nu$, whose relevant
form factors are discussed in the next Section. There theoretical and
experimental errors are of comparable size.

QED and radiative corrections to leptonic decays of $B$ mesons are of interest
as they may be enhanced. Refs.~\cite{Beneke:2017vpq,Beneke:2019slt} discuss
terms of O($m_b/\Lambda)$ and logarithmic enhancements, for $B_{(s)} \to
\mu^+\mu^-$. In general one expects large collinear logarithms of the form
$\log\left(m_b/m_\ell\right)$ because of the different scales involved.  It
would obviously be very important to confirm such effects in a fully
non-perturbative lattice computation. The situation is not as advanced as for
pions and kaons where the strategy put forward in Ref.~\cite{Carrasco:2015xwa}
could be used. This method relies on the use of a point-like approximation in
part of the computation (the real photon emission), which is equivalent to
neglecting structure dependent contributions. Such contributions are large for
heavy mesons as shown in
Refs.~\cite{Desiderio:2020oej,Giusti:2023pot,Frezzotti:2023ygt}, where radiative
decays of $D$ mesons have been studied on the lattice.  The effect should be
even more pronounced for $B$ mesons in which case one expects a faster
deterioration of the noise to signal ratio for the relevant correlation
functions~\cite{Frezzotti:2023ygt} in Euclidean time.
\section{Exclusive semileptonic decays \label{sec:semi}}
Differential decay rates for semileptonic decays of a heavy-hadron $H$ into a
lighter hadron $h$ are commonly parameterised as a sum of products of
non-perturbative form factors $f_X$ and known kinematic factors $\mathcal{K}_X$,
\begin{equation}
  \frac{\mathrm{d}\Gamma (H \to h)}{\mathrm{d}q^2} = \sum_X \mathcal{K}_X f_X(q^2)\,,
\end{equation}
where $q^\mu = p_H^\mu - p^\mu_h$ is the momentum transfer onto the lepton
pair. Depending on whether the process is tree-level or loop-induced, whether
the final state is a pseudoscalar~(PS) or a vector~(V) state, and whether it is
a baryonic or a mesonic decay the number of form factors $X$ that are summed
over differs. For the simplest case of a mesonic tree-level PS~$\to$~PS
transition (e.g. $B \to \pi \ell \nu$) there are two form factors ($f_+, f_0$);
for a loop-level induced PS~$\to$~PS transition (e.g. $B \to K\ell^+\ell^-$)
there are three form factors ($f_+, f_0, f_T$). For tree-level induced
PS~$\to$~V decays (e.g. $B \to D^*\ell\nu$) there are four ($V, A_0, A_1,
A_2$)\footnote{These results are often quoted in the helicity basis
  $f$,$g$,$\mathcal{F}_1$ and $\mathcal{F}_2$.} and for loop-level induced
PS~$\to$~V (e.g. $B \to K^* \ell \nu$) there are seven ($V, A_0, A_1, A_2, T_1,
T_2, T_3$). Depending on the decay, there are also kinematic constraints,
relating form factors at particular values of $q^2$ to each other.

\subsection{Challenges for form factors}
The data analysis relating simulated correlation functions to physical form
factors is lengthy and challenging. Whilst it is in principle understood how to
perform the simulations required for the prediction of semileptonic form
factors, it is costly to generate data sets which enable data-driven control
(guided by theory) over the required extrapolations listed in the following:

\begin{description}
\item[\underline{Excited states:}] On the lattice, form factors can be related
  to Euclidean matrix elements $\matrixel{h}{J}{H}$ of some current $J$ between
  the initial $H$ and final state $h$ and are extracted as the ground state
  matrix-element of a three-point function $C_3^{H\to h}(t;\Delta T) =
  \avg{O_h(\Delta T) J(t) O_H^\dagger(0)}$. Accurately isolating the correct
  matrix element is a trade-off between sufficiently large separations $\Delta T
  \gg t \gg 0$ allowing for the excited state contributions to be exponentially
  suppressed, and small enough values of $\Delta T$ to maintain good control
  over statistical uncertainties. Based on chiral perturbation theory
  ($\chi$PT), Ref.~\cite{Bar:2023sef} recently advocated that the expected first
  excited state energy to this type of three-point functions is expected to be
  of the size $m_B + m_\pi$ and might therefore potentially be hard to
  disentangle from the desired signal.  As an example, JLQCD in their
  computation of form factors for $B \to D^*$ semileptonic
  transitions~\cite{Aoki:2023qpa}, which we discuss in detail later, shows
  results for the three-point functions for different source-sink separations,
  concluding that results stabilise for a separation of about 1.5 fm.

\item[\underline{(Heavy-quark)-chiral-continuum extrapolation:}] Typical
  calculations of these form factors either take place at the physical $b$-quark
  mass using an effective action inspired discretisation, or at
  lighter-than-physical heavy-quark masses. All of the results currently
  available in the literature utilise ensembles with heavier-than-physical
  light-quark masses, so that a chiral extrapolation is required in addition to
  the continuum limit. These extrapolations are typically guided by heavy meson
  $\chi$PT (HM$\chi$PT) and therefore rely on inputs stemming from the world at
  physical quark masses and on low energy constants (LECs), which have to be
  determined. This makes controlled predictions very challenging when
  simulations take place at heavier-than-physical light-quark masses and
  lighter-than-physical heavy-quark masses.

\item[\underline{Kinematic coverage:}] Accessible Fourier momenta are of the form
  $a\vec{p} = \vec{n}2 \pi a/L $ where $\vec{n}$ is a vector of integers and
  $\abs{a\vec{p}}$ is required to remain small in order to control
  discretisation effects. Due to the heavy $b$-quark mass, it is not typically
  possible to cover the kinematically allowed range $q^2 \in [0, (m_H-m_h)^2
    \equiv q^2_\mathrm{max}]$ whilst maintaining control over discretisation
  effects. As a consequence, current calculations only cover a portion
  $[q^2_\mathrm{min,dat},q^2_\mathrm{max}]$ of the kinematic range at physical
  kinematics.\footnote{At unphysical kinematics, typically for $m_h \ll m_b$,
    some calculations cover the kinematic range corresponding to the choice of
    simulated masses.} This necessitates extrapolations of the lattice form
  factors over the full kinematical range.
\end{description}

\subsection{$z$-expansions and unitarity constraints}
The extrapolations over the full kinematic range are typically carried out as a
model-independent \emph{$z$-expansion}, a conformal mapping based on
Ref.~\cite{Boyd:1994tt} (or variants
thereof~\cite{Bourrely:2008za,Caprini:1997mu}). After removing terms
corresponding to sub-threshold poles ($P_X$) and an appropriate normalisation
($\phi_X(q^2)$), the form factor is expressed as a polynomial in the conformal
variable $z$,
\begin{equation}
  P_X(q^2) \phi_X(q^2) f_X(q^2) = \sum_{n=0}^\infty a_n z^n.
\end{equation}
Since the conformal variable satisfies $\abs{z} < 1$, this is a convergent
series that, for a given truncation, can then be fitted to available lattice
data. Based on dispersion relations, a unitarity bound $\sum \abs{a_n}^2 \leq 1$
can be derived.\footnote{Depending on the pole structure of the decay, the exact
  form of this bound can vary~\cite{Gubernari:2020eft, Flynn:2023qmi}.} Whether
the unitarity bound is satisfied can either be checked \emph{a~posteriori} or
the constraint can be imposed directly in the fit, either via the dispersive
matrix method~\cite{DiCarlo:2021dzg} or via a recently proposed Bayesian
Inference framework~\cite{Flynn:2023qmi}. The latter allows to trade truncation
effects for statistical noise encoding our ignorance of the neglected higher
order terms, which are however regulated by the unitarity constraint. In
general, multiple decay channels with the same quark-level transition can be
incorporated into the same unitarity constraint, strengthening the constraints
on $a_n$ coefficients.  As an example, since the current in $B \to \pi \ell \nu$
and in $B_s \to K \ell \nu$ transitions is the same, the unitarity constraint
involves a combination of the corresponding form factors, which should therefore
be fitted simultaneously.  The unitarity bound will be closer to saturation
(compared to imposing the constraint on the individual decay channels),
resulting in less freedom and hence tighter constraints on the higher order
$z$-expansion coefficients. This can be systematically improved by including
further decays of the same current (e.g. $\Lambda_b \to p \ell \nu$, $B_s \to
K^* \ell \nu$, ...).

\subsection{Analysis strategies}
Most collaborations perform their analyses in a multi-step procedure, consisting of
\begin{enumerate}
\item The extraction of the desired masses and hadronic matrix elements from the
  correlation function data.
\item The extrapolation to zero lattice spacing and physical quark masses and a
  continuous description of the momentum transfer in the kinematic range covered
  by the data. The fit is then evaluated at a number of \emph{reference
    $q^2$-values} and a full error budget including the correlations between the
  form factors at the reference values is assembled.
\item An extrapolation over the full kinematic range, typically using a
  model-independent $z$-extrapolation.
\end{enumerate}
Some collaborations have advocated combining the second and the third step into
the \emph{modified $z$-expansion}~\cite{Na:2010uf} in which the coefficients of
the $z$-expansion are changed into functions of the lattice spacing and the
quark masses. However, since this is no longer based on an underlying effective
field theory, this approach might introduce model dependences, which are hard to
quantify with current data.

In the following, we restrict the discussion to form factor calculations with
results of multiple collaborations available and cases where tensions have been
observed. In particular we will address $B_{(s)} \to$~PS and $B_{(s)} \to$~V
tree-level decays that can be used for the determination of CKM matrix elements
and which tend to be most mature.
\begin{figure*}
  \includegraphics[width=0.47\textwidth]{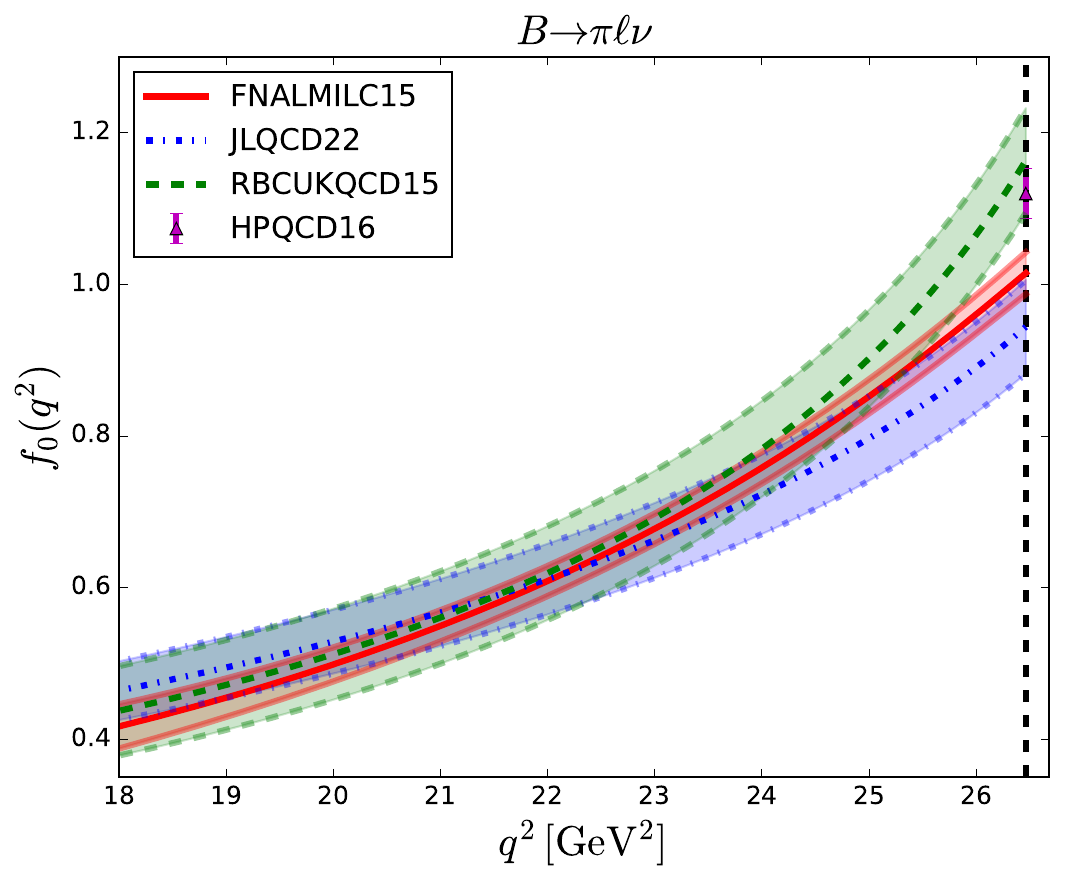}
  \includegraphics[width=0.47\textwidth]{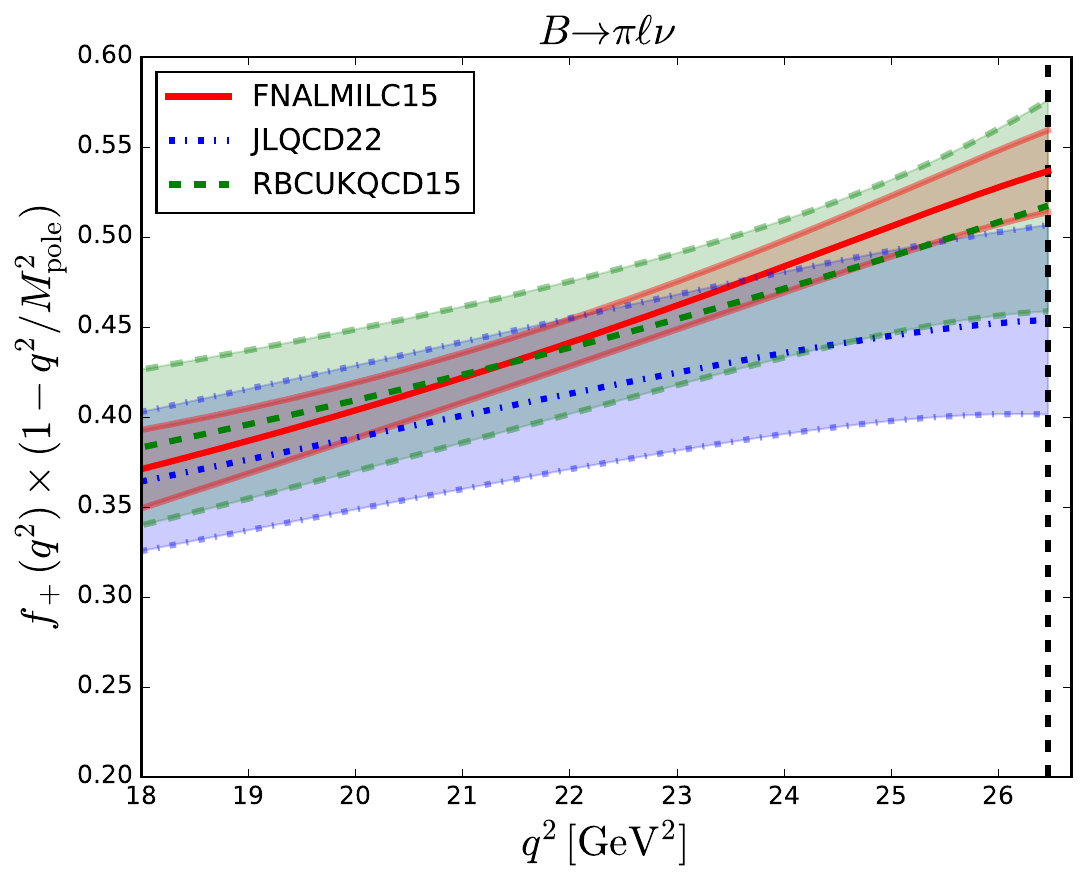}
  \caption{Comparison plot of recent results for $B\to\pi$ form factors from
    Refs.~\cite{Colquhoun:2022atw,Flynn:2015mha,FermilabLattice:2015mwy}. The
    individual data point HPQCD16~\cite{Colquhoun:2015mfa} stems from a
    $N_f=2+1+1$ calculation, but only provides as results
    $f_0(q^2_\mathrm{max})$.}
  \label{fig:semi_b_to_u}
\end{figure*}
\subsection{Tree-level $B_{(s)}\to$~PS decays}
Results for $B \to \pi \ell \nu$ form factors covering part of the kinematic
range exist from JLQCD~\cite{Colquhoun:2022atw} (JLQCD22),
RBC/UKQCD~\cite{Flynn:2015mha} (RBCUKQC15) and
Fermilab/MILC~\cite{FermilabLattice:2015mwy} (FNALMILC15).\footnote{The result
  by HPQCD~\cite{Dalgic:2006dt} uses gauge field configurations which are also
  part of FNAL/MILC15 calculation but span a smaller range in the key parameters
  such as the lattice spacing and pion masses.}  For $B_s\to K \ell \nu$,
results from RBC/UKQCD~\cite{Flynn:2023nhi} (superseding
Ref.~\cite{Flynn:2015mha}), FNAL/MILC~\cite{FermilabLattice:2019ikx}, and
HPQCD~\cite{Bouchard:2014ypa} exist. Ref.~\cite{Colquhoun:2022atw} features
domain wall fermions for all quarks and extrapolates from $m_c \leq m_h \lesssim
2.44m_c$ to the physical $b$-quark mass. All other results utilise effective
action approaches for the $b$ quark.
Figure~\ref{fig:semi_b_to_u} summarises the status for $B \to \pi$, in the range
where lattice data typically exist. One can clearly see a tension near
$q^2_\mathrm{max}$ (the vertical black dashed line) for the $f_0$ form
factor. This tension is more severe than is apparent from a plot, since the
reference values underlying the individual datasets are highly correlated. When
attempting to jointly fit synthetic data points for these results, the FLAG
collaboration reports
$\chi^2/\mathrm{dof}=43.6/12$~\cite{FlavourLatticeAveragingGroupFLAG:2021npn},
necessitating an error inflation by the PDG scale factor
$\sqrt{\chi^2/\mathrm{dof}}$. The situation is similar for $B_s \to K \ell \nu$
as is demonstrated in Ref.~\cite{Flynn:2023qmi}\footnote{The FLAG report does
  not yet include the recent RBC/UKQCD23~\cite{Flynn:2023nhi} result.}, where
the world data cannot be jointly fitted with an acceptable $p$-value and
$\chi^2/\mathrm{dof}=3.89$ is the best value that can be achieved. A possible
explanation for some of the discrepancies between results, stemming from the
assignment of external parameters to the pole structure imposed in the
HM$\chi$PT has been put forward in Ref.~\cite{Flynn:2023qmi}, but further
independent computations are needed to confirm this.

For the $b\to c$ transitions $B_{(s)} \to D_{(s)}$ there are fewer independent
results. $B \to D$ form factors have been published by HPQCD~\cite{Na:2015kha}
and FNAL/MILC~\cite{MILC:2015uhg}. The calculations based on two and four
lattice spacings respectively are compatible. The FNAL/MILC result dominates the
combined fit, but due to overlapping gauge field configurations (the $N_f=2+1$
asqtad ensembles) between the two computations these results are expected to be
statistically correlated.

Two results by the HPQCD collaboration~\cite{Monahan:2017uby,McLean:2019qcx}
exist for $B_{s} \to D_s$ form factors. The former based on the $N_f=2+1$ asqtad
ensembles with NRQCD $b$-quark and HISQ $c$-quark, the latter based on the
$N_f=2+1+1$ HISQ ensembles with HISQ for all quark flavours. Whilst the result
covers the entire kinematic range at unphysical kinematics, this is not the case
at the physical values of the $b$-quark mass. To predict form factors at
physical masses, an extrapolation of the form factors in the heavy-quark mass is
required. This is complicated since the momentum transfer $q^2$ is also a
function of the heavy-quark mass. Ref.~\cite{McLean:2019qcx} uses a modified
$z$-expansion to simultaneously perform the extrapolations to physical light and
heavy-quark masses, zero lattice spacing and a continuous description of the
momentum transfer $q^2$ in the physical range.

Additional independent results for all of the above would be very desirable in
order to address tensions (where present) and to test the choices that have been
made in the literature. Computations are ongoing by several collaborations, in
particular by the RBC/UKQCD~\cite{Flynn:2019jbg,Flynn:2021ttz}, the JLQCD
collaboration~\cite{Kaneko:2021tlw} and the FNAL/MILC
collaboration~\cite{Lytle:2023xuq}.

\subsection{$B_{(s)} \to D^{*}_{(s)}$ decays}
In recent years the process $B\to D^{*}\ell \nu$ (as well as $B\to D \ell \nu$ covered in the previous section) has received much attention,
due to long-standing tensions between experiment and theory concerning these
decays~\cite{HFLAV:2022pwe}. In 2021 FNAL/MILC~\cite{FermilabLattice:2021cdg}
produced the first comprehensive calculation of the four form factors involved
away from the zero-recoil point. In 2023 two additional results appeared
(HPQCD~\cite{Harrison:2023dzh} and JLQCD~\cite{Aoki:2023qpa}), which, at the
time of writing this report, are only available as pre-prints. Due to the
interest in these very recent results, we compare the three calculations in
detail in the following

\begin{description}
\item[\underline{Lattice set-ups:}] The FNAL/MILC~\cite{FermilabLattice:2021cdg}
  work uses 15 $N_f=2+1$ ensembles with the asqtad fermion action with five
  lattice spacings $a \in [0.045,0.15]\,\mathrm{fm}$ and a range of
  (root-mean-square) pion masses\footnote{For staggered quarks several light
    states have the quantum numbers of the pion. Their masses differ by {\it
      taste-breaking} $O(a^2)$ effects.  The lightest state is called
    Goldstone-pion.} down to $m_\pi^{RMS} \sim 250\,\mathrm{MeV}$. Both $b$ and
  $c$ quarks are simulated with the Fermilab action. The
  HPQCD~\cite{Harrison:2023dzh} computation uses five $N_f=2+1+1$ ensembles with
  the HISQ action for all sea and valence quarks including three lattice
  spacings $a\in [0.044,0.090]$ and two ensembles with physical Goldstone-pion
  masses
  The JLQCD~\cite{Aoki:2023qpa} calculation uses nine
  $N_f=2+1$ ensembles with the domain wall action for all quarks and three
  lattice spacings $a \in [0.04, 0.08] \,\mathrm{fm}$. Simulations take place in
  the range $226\,\mathrm{MeV} \lesssim m_\pi \lesssim 500\,\mathrm{MeV}$. HPQCD
  and JLQCD simulate a range of heavy-quark masses below the physical $b$-quark mass
  with the constraint $am_q \leq 0.7$ and $0.8$, respectively and therefore
  require an extrapolation in the heavy-quark mass.

  Heavy-to-heavy transitions are often described as a function of the kinematic
  variable $w = v_{B} \cdot v_{D^*}$. Since all the mentioned works take
  place in the $B$-meson rest-frame this reduces to $w =
  E_{D^*}/M_{D^*}$. The range of this parameter approximately
  corresponding to the semileptonic range is $w\in[1,1.5]$. The FNAL/MILC, and
  JLQCD results cover the range from $w=1$ to $w\sim 1.175$ and $1.1$,
  respectively on all their ensembles. The HPQCD result covers the range from
  $w=1$ to 1.05, 1.20, and 1.39 on their $a\approx 0.09, 0.06$, and
  $0.044\,\mathrm{fm}$ ensembles, respectively.

\item[\underline{Lattice to continuum:}] Each of the three works perform a fit
  to simultaneously extrapolate to the continuum and to physical quark masses
  and to obtain a continuous description of the form factors as a function of
  $w$.

  All three collaborations use $\chi$PT (or staggered variants of this) to
  extrapolate to the physical pion mass and describe the kinematic behaviour as
  an expansion in $(w-1)^k$ up to quadratic (JLQCD, FNAL/MILC) or cubic (HPQCD)
  order.

  FNAL/MILC includes discretisation effects of order $\alpha_s a\Lambda$,
  $(a\Lambda)^2$ and $(a\Lambda)^3$. JLQCD accounts for terms $(a\Lambda)^2$ and
  $(am_q)^2$ since due to the use of domain wall fermions discretisation effects
  from odd powers of the lattice spacing are absent (up to terms proportional to
  the residual mass which are negligible). JLQCD's heavy quark extrapolation is
  performed by first dividing out the matching factor between HQET and QCD and
  then parameterising this result in powers of $1/m_h$ up to order 1. HPQCD
  parameterises discretisation effects and the heavy-quark mass dependence
  as products of $(am_c)^{2i}(am_h)^{2j} [(\Lambda/M_{H_s})^k -
    (\Lambda/M_{B_s})^k]$ for $i,j,k=0,1,2,3$. One such term is present for each
  order of $(w-1)^n$ for $n=0,1,2,3$.

  Per form factor, the resulting fits typically have 8 parameters for JLQCD, 10
  parameters for FNAL/MILC and more than 250 parameters for HPQCD. JLQCD
  performs the fits as frequentist $\chi^2$-minimisations, whilst FNAL/MILC and
  HPQCD use a Bayesian framework with Gaussian priors for their fit parameters.

  One caveat concerning the HPQCD computation~\cite{Harrison:2023dzh} is that
  for each form factor the $B \to D^*$ and $B_s \to D_s^*$ data are fitted
  jointly, imposing that all the fit parameters are the same. The only freedom
  for the fit to distinguish between $B \to D^*$ and $B_s\to D_s^*$ is by
  some of these parameters multiplying an SU(3) breaking term. Since
  Ref.~\cite{Harrison:2023dzh} does not present fits to the individual channels
  it is not possible to quantify the impact of this
  choice. Ref.~\cite{Harrison:2023dzh} however compares the result obtained for
  $B_s \to D_s^*$ from the simultaneous fit to $B \to D^\star$ and $B_s \to
  D_s^\star $~\cite{Harrison:2023dzh} to the previous computation of only $B_s
  \to D_s^*$ form factors~\cite{Harrison:2021tol} in the helicity basis,
  which will be described below.
\item[\underline{$z$-expansions:}] Fermilab/MILC and JLQCD perform a two-step
  analysis, evaluating their chiral-continuum-heavy-quark extrapolation at three
  values of $w$ in the range of data where the simulations took place
  ($w_\mathrm{ref}^\mathrm{FNAL/MILC}\in\{1.03,1.10,1.17\}$,
  $w_\mathrm{ref}^\mathrm{JLQCD} \in\{ 1.025, 1.060, 1.100\}$). They each
  assemble a full error budget for all four form factors at these data points
  and quantify all correlations. These reference values serve as inputs for a
  subsequent $z$-expansion, which extrapolates the form factors over the full
  kinematic range.

  HPQCD provides results for a $z$ expansion, but does not clarify what input
  data was used for this $z$-expansion. It is unclear whether the fit was
  performed to reference $w$-values, and if so, how they were chosen, how the
  full error budget for these reference values is computed and what the
  correlations between these values are.

  Since all three collaborations use the same conventions for the $z$-expansion,
  the results can be directly compared and the coefficients from the three
  collaborations are listed in Table~\ref{tab:BtoDstar_zexp} and displayed in
  Figure~\ref{fig:BtoDstarcoefs}.\footnote{One minor difference is however, that
    JLQCD enforces the kinematic constraint relating $\mathcal{F}_1$ and
    $\mathcal{F}_2$ at $w_\mathrm{max}$, whilst the other two collaborations
    check it \emph{a posteriori}.} There are clearly several
  tensions between the results for the lower order coefficients which will need
  to be understood.

  \begin{figure}
    \includegraphics[width=.48\textwidth]{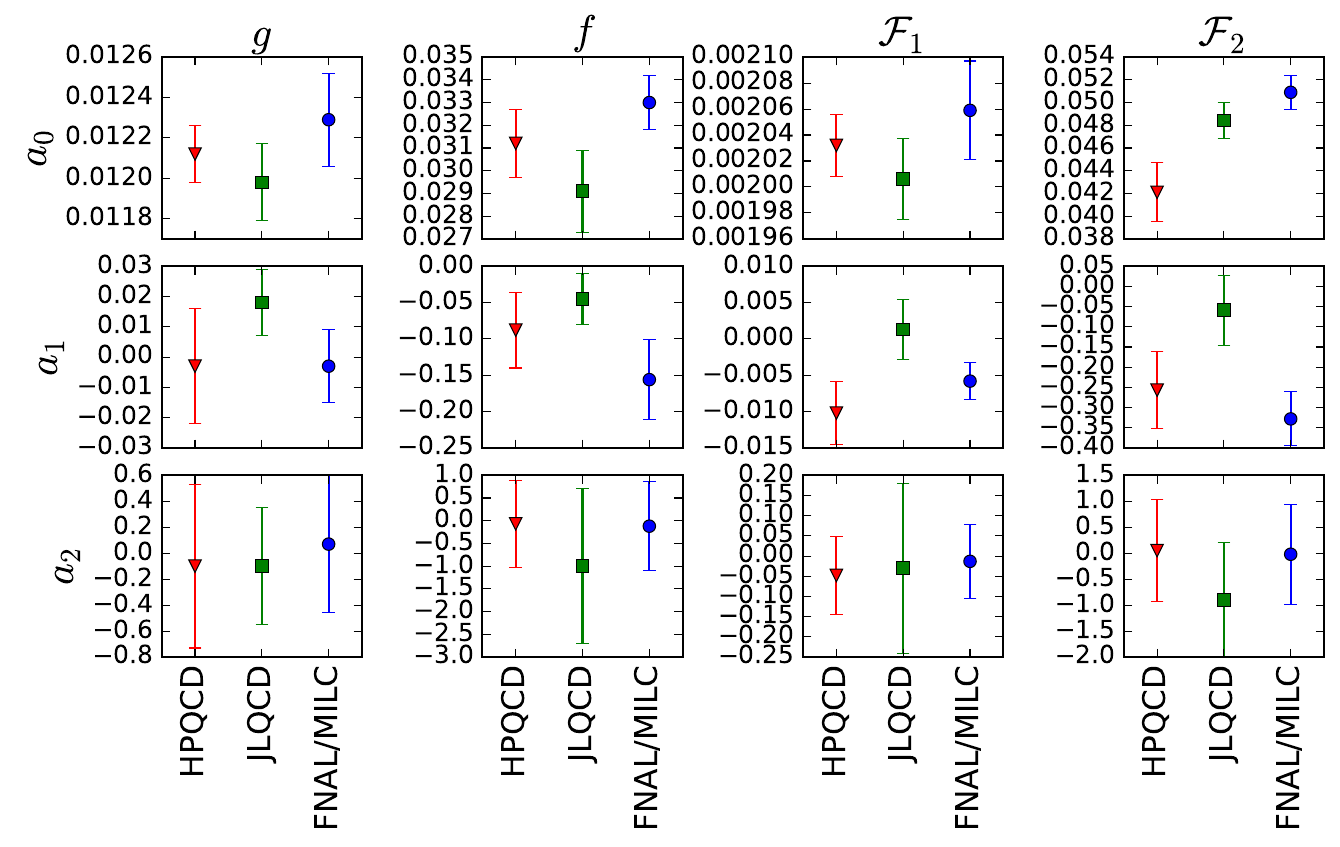}
    \caption{Comparison of the $z$-expansion coefficients for the $B \to D^*$
      form the factors $g$, $f$, $\mathcal{F}_1$ and $\mathcal{F}_2$ obtained by
      FNAL/MILC (blue circles), HPQCD (red triangles), and JLQCD (green
      squares).}
    \label{fig:BtoDstarcoefs}
    \end{figure}
\item[\underline{$B_s \to D_s^*$ form factor results by HPQCD:}] We now turn
  our attention to the comparison of the results for the $B_s\to D_s^*$ form
  factors from the two HPQCD calculations from 2023~\cite{Harrison:2023dzh} and
  2021~\cite{Harrison:2021tol}.

  The only difference in the ensembles underlying this dataset is one additional
  ensemble in~\cite{Harrison:2023dzh} at the intermediate lattice spacing with
  physical pion mass providing additional data in the range $w \in
  [1.0,1.2]$. Since there are no valence light quarks in the decay $B_s \to
  D_s^*$ this is not expected to significantly alter the result. Furthermore,
  stemming from largely the same ensembles, the two results are highly
  statistically correlated, so that any discrepancies between the results are
  expected to stem from systematic uncertainties.

The form factors $f$ and $g$ largely remain within one standard deviation, but
with a reduction of the quoted uncertainties by a factor of approximately 3.5. A
similar reduction of uncertainties is quoted for the form factor
$\mathcal{F}_1$. Whilst this form factor agrees with the previous result near
$w\sim 1$, the slope with $w$ is very different and in the intermediate $w$
region, there is an approximately $2 \sigma$ tension between the two results.
For the final form factor $\mathcal{F}_2$ the error reduction is less
significant, but the slope near $w\approx 1$ is different between the two
calculations.

The authors of Ref.~\cite{Harrison:2023dzh} explain the observed differences
with the change in analysis strategy from an expansion in powers of the
conformal variable $z$ to an expansion in the variable $(w-1)$ and by the joint
fitting of $B\to D^*$ and $B_s \to D_s^*$ data. Since the joint fits in
Ref.~\cite{Harrison:2023dzh} are dominated by the statistically more precise
$B_s \to D_s^*$ data, the latter seems unlikely to explain the
differences. According to Ref.~\cite{Harrison:2023dzh}, the main contribution to
the total uncertainty is statistical, with systematic uncertainties accounting
for at most 15\%, 10\%, 25\%, and 20\% of the variance of $f$, $g$,
$\mathcal{F}_1$, and $\mathcal{F}_2$, respectively. Further
investigations are required to understand the differences between
Refs.~\cite{Harrison:2023dzh} and~\cite{Harrison:2021tol}. It would be
interesting if correlated differences between the two results could be produced.
\end{description}

\begin{table}
  \caption{$z$-expansion coefficients quoted by the three collaborations.}
  \begin{tabular}{ c lll }\hline
    Coeffs & HPQCD & JLQCD & FNAL/MILC \\\hline
    $a_0^{\g}$ & \phantom{$-$}0.0312(15)   &\phantom{$-$}0.0291(18)  & \phantom{$-$}0.0330(12)\\
    $a_1^{\g}$ & $-$0.088(52)    &$-$0.045(35)   & $-$0.156(55) \\
    $a_2^{\g}$ & $-$0.07(95)     &$-$1.0(1.7)    & $-$0.12(98)\vspace{.05cm}\\\hline
    $a_0^{\f}$ & \phantom{$-$}0.01212(14)  &\phantom{$-$}0.01198(19) & \phantom{$-$}0.01229(23)\\
    $a_1^{\f}$ & $-$0.003(19)    &\phantom{$-$}0.018(11)   & $-$0.003(12)\\
    $a_2^{\f}$ & $-$0.10(63)     &$-$0.10(45)    & \phantom{$-$}0.07(53)\vspace{.05cm}\\\hline
    \vspace{.05cm}$a_0^{\Fo}$ & \phantom{$-$}0.002032(24)&\phantom{$-$}0.002006(31)& \phantom{$-$}0.002059(38)\\
    $a_1^{\Fo}$ & $-$0.0102(43)  &\phantom{$-$}0.0013(41)  & $-$0.0058(25)\\
    $a_2^{\Fo}$ & $-$0.048(96)   &$-$0.03(21)    & $-$0.013(91)\vspace{.05cm}\\\hline
    \vspace{.05cm}$a_0^{\Ft}$ & \phantom{$-$}0.0421(26)  &\phantom{$-$}0.0484(16)  & \phantom{$-$}0.0509(15)\\
    $a_1^{\Ft}$ & $-$0.257(95)   &$-$0.059(87)   & $-$0.328(67)\\
    $a_2^{\Ft}$ & \phantom{$-$}0.05(98)    &$-$0.9(1.1)    & $-$0.02(96)\vspace{.05cm}\\\hline
  \end{tabular}
  \label{tab:BtoDstar_zexp}
\end{table}

\subsection{Desirable benchmark quantities, comparisons and checks \label{subsec:benchmarks}}
The extrapolations required to predict form factors that can be related to the
physical world from the underlying lattice data points are more complex than
those for quantities that do not depend on a kinematic variable. This is mainly
due to the fact that the kinematically allowed range ($q^2_\mathrm{max}$ or
equivalently $w_\mathrm{max}$) changes as a function of the mass of the initial
and the final states. This is further aggravated by the increase of statistical
noise as the heavy-quark mass becomes heavier and as larger momenta are
induced. In particular when simulating multiple heavy-quark masses, this often
leads to a comparably small number of precise data points (at small $m_h$, small
$\vec{p}$) and a lot of data points with sometimes orders of magnitude larger
uncertainties (large $m_h$, large $\vec{p}$). These in turn need to be described
by (often complicated) multidimensional fits. As a result a small portion of
data points drive the fit, typically those furthest away from the desired
physical parameters. The fit functions that are employed often have many
parameters. In order to assess the weight of the different data in the fit and
to determine the relevant fit parameters, it could help to $i)$ fit only the
less precise data, to see what their effect is, and/or $ii)$ start by only
including the most precise data (and therefore using simple fit ans{\"a}tze) and
then adding less precise data until results stabilise.
This would provide an assessment of which portions of the covered parameter space
constrain the fit.  On a related note we remark that whilst datapoints with very
large uncertainties (relative to the majority of the data) do not significantly
contribute to the $\chi^2$, they alter the interpretation of how good a
particular fit result is, by reducing the $\chi^2/\mathrm{dof}$ without adding
much information. In the extreme case where the relative uncertainty of data
points differs by orders of magnitude, it might be worth to develop a notion of
\emph{effective degrees of freedom}.

Given the tensions in several form factor computations, it would be desirable to
devise some easier-to-compute benchmark quantities that all collaborations could
provide, somewhat analogous to the window quantities in the hadronic vacuum
polarisation contribution to the anomalous magnetic moment of the
muon~\cite{RBC:2018dos}. These should be designed to allow for comparisons
between different computations with reduced sources of systematic
uncertainties. Particularly well suited are quantities which disentangle the
effects of the different dimensions of the fit, such as the chiral, continuum,
kinematic and heavy quark extrapolations. We conclude this section with the
incomplete list of suggested benchmark quantities below:
\begin{itemize}
\item A full error budget for the form factors $f_0(q^2_\mathrm{max})$ for
  PS~$\to$~PS transitions and $f(w=1)$ for PS~$\to$~V transitions solely based
  on the zero momentum data points. Since all hadrons are at rest, these can be
  obtained from the most precise data points and no interpolation in the final
  state energy is required.
\item Generalising the previous point: the form factor at kinematic reference
  points. Whilst several collaborations tend to provide these as input for
  subsequent kinematic extrapolations, they are typically obtained from a joint
  fit of all simulated momentum values. Performing such checks only from data in
  the vicinity of the kinematic reference point would allow more direct
  comparisons which do not rely on the broader choice of kinematic description
  of the data. For example, differences between choosing a modified $z$
  expansion, an expansion in $(w-1)$ or an expansion in the final state energy
  could be assessed. It would also shed light on how much of the information
  content stems from simulated data near the kinematic reference point as
  opposed to potentially more precise data which is kinematically far away.
  
\item For simulations which take place at lighter-than-physical heavy-quark
  masses, it would be valuable to perform the chiral-continuum-kinematic
  extrapolation at fixed (unphysical) heavy-quark mass. If data was generated
  with this in mind, one could simulate data at, for example, half the $b$-quark
  mass and make predictions of the form factors at this choice of
  kinematics. This would remove the heavy-quark extrapolation dimension from the
  fit and hence disentangle the two most complicated extrapolations: the
  heavy-quark mass extrapolation and the continuum-limit. Furthermore, since the
  simulated heavy-quark mass in lattice units is smaller, better control over
  the continuum limit would be expected. These results could be compared between
  different collaborations.

\item Assessment of the contamination of ground state matrix elements and masses
  by excited states is a recurring feature of lattice QCD calculations. The
  excited state energies obtained from correlation function fits are physical
  quantities and should (up to discretisation effects) agree for different
  fermion formulations at equivalent quark masses. If this data was included in
  publications, it would help to build confidence in the absence of excited
  state contamination or to understand their nature.
\item In connection with the discussion on the varying size of statistical
  uncertainties, it would be valuable if the statistical correlations between
  all data points stemming from the same ensemble were given. Since this is a
  vital ingredient of the fit extrapolating to physical parameters, this
  information is required in order to reproduce results. Furthermore, these
  correlations should be universal (up to discretisation effects), a fact that
  could be checked if these results were provided by all collaborations.
\end{itemize}

\section{Neutral meson mixing \label{sec:mix}}
\begin{figure}
  \includegraphics[width=.45\textwidth]{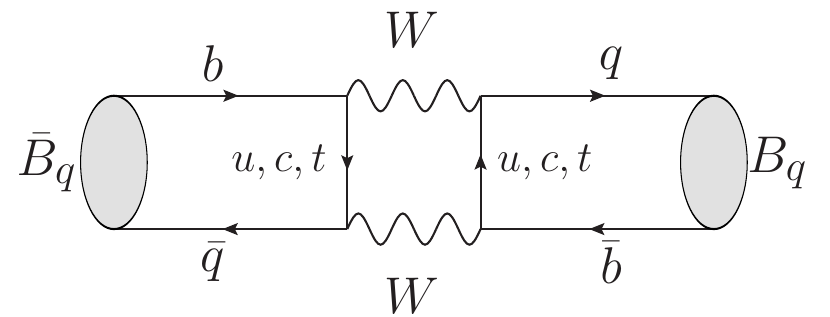}
  \caption{Example box diagram mediating neutral meson mixing}
  \label{fig:box}
\end{figure}

\begin{table*}
  \caption{Some key properties of the results discussed in the text. The
    abbreviations for the various fermion actions stand for Wilson twisted mass
    (Wtm), $a^2$-tadpole improved staggered quarks (asqtad), highly improved
    staggered quarks (HISQ), domain wall fermions (DWF) and Osterwalder-Seiler
    (OS).
    $^\dagger$The root-mean-square pion mass is listed. The
    corresponding goldstone pion masses ranges are $[177,555]$ for FNAL/MILC and
    $[131,313]$ for HPQCD.}
  \begin{tabular}{l|cccc}
    Work & ETM~\cite{ETM:2013jap} & FNAL/MILC~\cite{FermilabLattice:2016ipl} & HPQCD~\cite{Dowdall:2019bea} & RBC/UKQCD~\cite{Boyle:2018knm}\\\hline
    $N_f$ & 2 & 2+1 & 2+1+1 & 2+1\\
    sea action & Wtm & asqtad & HISQ & DWF \\
    valence light action & OS & asqtad & HISQ & DWF\\
    valence heavy action & OS & Fermilab & NRQCD & DWF\\
    physical heavy quark & ratio method & tuned & tuned & extrap. in $1/M_{H_s}$\\
    directly computed & $f$, $\mathcal{B}$ & $f\sqrt{\mathcal{B}}$ & $\mathcal{B}$ & $\xi$, $\mathcal{B}_s/\mathcal{B}_d$\\
    $m^\mathrm{simulated}_\pi \,[\mathrm{MeV}]$ & [280,500] & [257,670]$^\dagger$ & [241,311]$^\dagger$& [139,430] \\
    range of $a \,[\mathrm{fm}]$ & $[0.052,0.098]$ & [0.045,0.12] & [0.088,0.147] & [0.073,0.114]\\
    number of $a$ & 4 & 4 & 3 & 3 \\
    renormalisation & RI-MOM & 1-loop PT & 1-loop PT & RI-SMOM \\
    continuum like mixing? & Yes & No & No & Yes \\
  \end{tabular}
  \label{tab:mixing}
\end{table*}

Neutral mesons such as the $B^0_{q}$ ($q=d,s$) mix with their antiparticles
$\bar{B}^0_{q}$ through box diagrams such as the one displayed in
Figure~\ref{fig:box}. Due to this mixing, the mass and flavour eigenstates of
the $B_q-\bar{B}_q$ system do not coincide, resulting in experimentally measurable
mass and width differences. Since the box diagrams are top-quark and therefore
short distance dominated, the relevant non-perturbative matrix elements are
calculable on the lattice. An operator product expansion of the box diagrams
above yields 5 independent parity-even dimension-6 operators, whose matrix
elements can be computed in LQCD.

For historical reasons, the matrix elements under consideration are typically
cast into the form of \emph{bag parameters} $\mathcal{B}_{B_q}^{(i)}$ where
$i=1,...,5$ and $q=d,s$, quantifying the departure of the matrix element from
the vacuum saturation approximation (VSA). These are defined as
\begin{equation}
  \mathcal{B}^{(i)}_{B_q}(\mu) = \frac{\matrixel{B_q}{\mathcal{O}_i(\mu)}{\bar{B}_q}}{f_{B_q}^2M_{B_q}^2 \eta_q^{(i)}(\mu)}\,,
\end{equation}
where the $\eta_q^{(i)}$ ensure that the expression is unity in the VSA. For
phenomenological applications, typically the quantities $f_{B_q}
\sqrt{\mathcal{B}_{B_q}^{(i)}}$ are of relevance, but depending on the
application, the bag parameters or the matrix elements themselves are also of
interest.

In the SM, the experimentally measured mass difference $\Delta m_q$ is related
to $f_{B_q}^2 \mathcal{B}^{(1)}_{B_q}$ by known multiplicative factors and the
product of CKM matrix elements $V_{tq}V_{tb}^*$ ($q=d,s$), and hence precise
knowledge of the non-perturbative inputs enables the determination of CKM matrix
element containing the top quark and thereby contribute to tests of CKM
unitarity. Since the non-perturbative matrix elements are independent of the
UV-properties of the theory under consideration they can also be used to probe
BSM models. Finally, several theoretical uncertainties cancel in the
SU(3)-breaking ratios $\mathcal{B}^{(1)}_{B_s}/\mathcal{B}^{(1)}_{B_d}$ and
$\xi$
\begin{equation}
  \xi \equiv \frac{f_{B_s} \mathcal{B}^{(1)}_{B_s}}{f_{B_d} \mathcal{B}^{(1)}_{B_d}} \propto \sqrt{\frac{\Delta m_s}{\Delta m_d}} \abs{\frac{V_{td}}{V_{ts}}}\,,
\end{equation}
so that non-perturbative determinations of this quantity provide more stringent
bounds on $\abs{V_{td}/V_{ts}}$.

Full results for all $5$ operators (and in both systems) are available from
ETM~\cite{ETM:2013jap} ($N_f=2$), FNAL/MILC~\cite{FermilabLattice:2016ipl}
($N_f=2+1$) and HPQCD~\cite{Dowdall:2019bea} ($N_f=2+1+1$). To date, the
RBC/UKQCD collaboration computed the ratios $f_{B_s}/f_B$,
$\mathcal{B}_{B_s}^{(1)}/\mathcal{B}_{B_d}^{(1)}$ and $\xi$ from $N_f=2+1$
QCD~\cite{Boyle:2018knm}. One difference between these computation is which
quantities are directly computed and which ones are inferred using available
results for $f_{B_q}$ from the literature. For maximally exploitable
phenomenological applications of these results, it would be desirable to have
access to $\mathcal{B}^{(i)}_{B_q}$ and $f_{B_q} \sqrt{\mathcal{B}_{B_q}^{(i)}}$
from the same computation, in order to be able to quantify all correlations
between these observables.

Table~\ref{tab:mixing} summarises some of the key properties of these results,
which are highly complementary: In addition to differences in the choice of sea,
valence light, and valence heavy quark actions, the range of lattice spacings
and simulated pion masses differs significantly.  When chiral symmetry is
maintained, the five bag parameters mix in a continuum-like fashion under
renormalisation, so the renormalisation pattern is block diagonal. In the
formulations used by the ETM and the RBC/UKQCD collaborations, this property is
preserved in the lattice simulation and hence simplifies the renormalisation
procedure. The treatment of the heavy-quark also differs, with FNAL/MILC and
HPQCD employing an effective action approach and ETM and RBC/UKQCD using a fully
relativistic set-up.

\begin{figure}
  \includegraphics[width=\columnwidth]{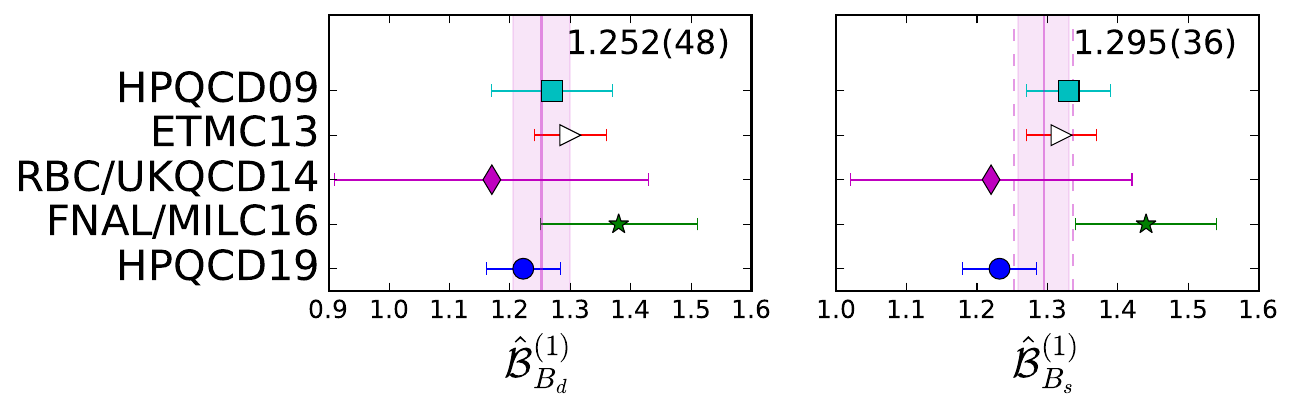}
  \includegraphics[width=\columnwidth]{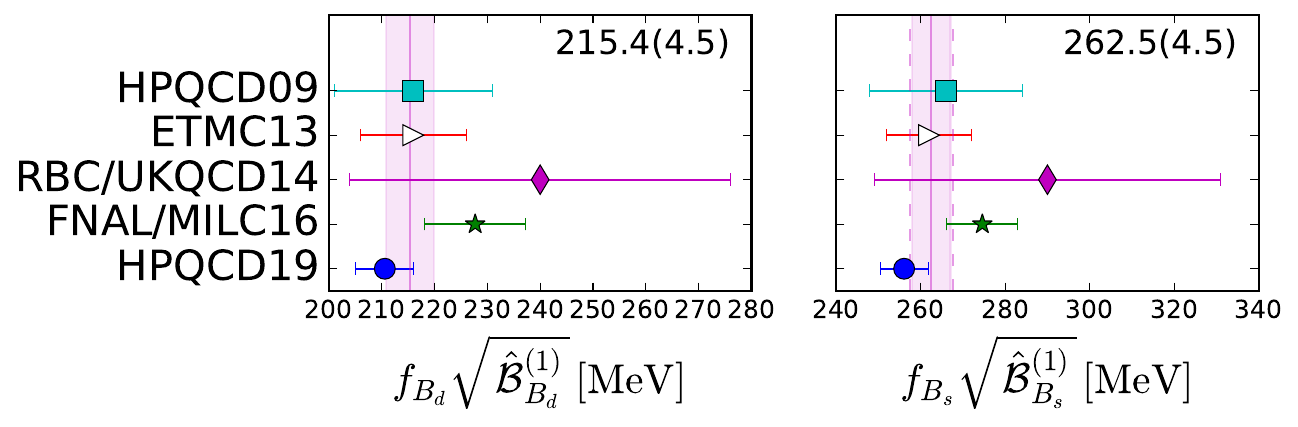}
  \includegraphics[width=\columnwidth]{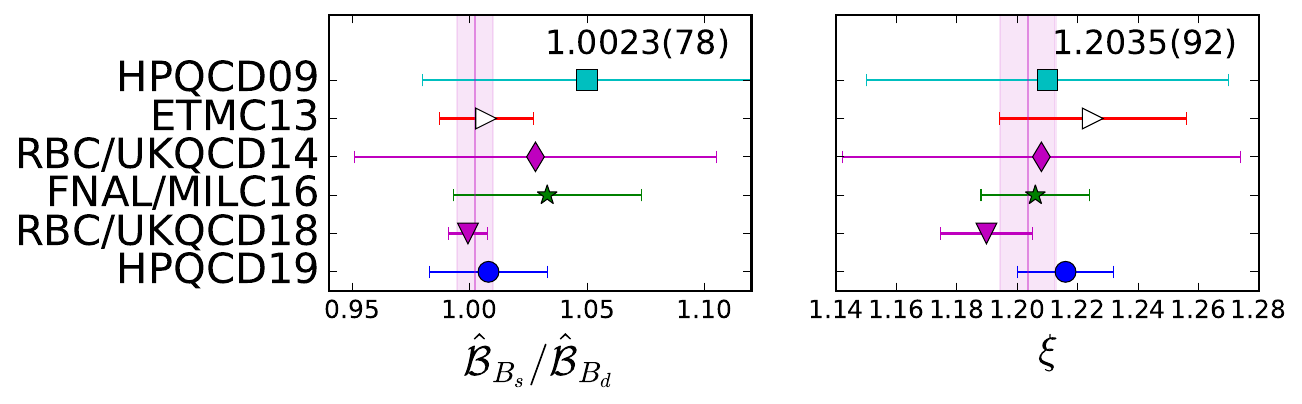}
  \caption{Summary of the first bag parameter and $f\sqrt{\mathcal{B}}$ in the
    RGI scheme. The magenta bands correspond to the values of a weighted average
    of $N_f >2$ results with its value given in the top right corner of each
    panel.}
  \label{fig:RGImixing}
\end{figure}
\begin{figure}
  \includegraphics[width=\columnwidth]{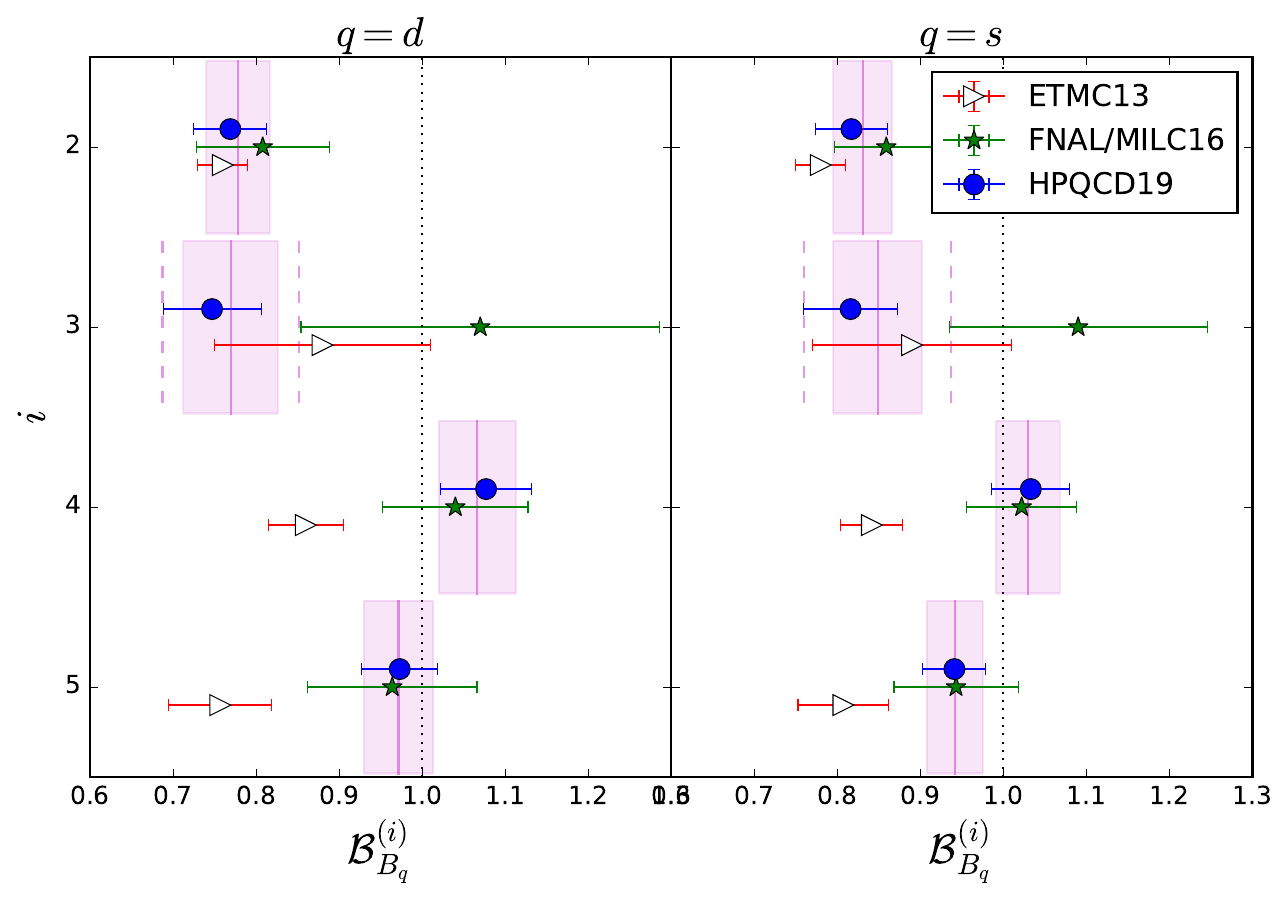}
  \caption{Using results from Refs.~\cite{ETM:2013jap},
    \cite{FermilabLattice:2016ipl} and \cite{Dowdall:2019bea} Results are quoted
    in $\overline{\mathrm{MS}}(m_b)$ and using the BBGLN scheme. Magenta bands
    are the weighted averages between $N_f=2+1$ and $N_f = 2+1+1$. In cases
    where $\sqrt{\chi^2/\mathrm{dof}}>1.25$ the dashed magenta lines indicate
    the uncertainty including the PDG scale factor of
    $\sqrt{\chi^2/\mathrm{dof}}$.}
  \label{fig:bags}
\end{figure}

Observables are often quoted in the Renormalisation Group Invariant (RGI)
scheme. Results for the phenomenologically relevant quantities
$\mathcal{\hat{B}}_{B_q}^{(1)}$, $f_{B_q} \sqrt{\mathcal{\hat{B}}_{B_q}^{(1)}}$
for $q=d,s$ and their SU(3)-breaking ratios are compared in
Fig.~\ref{fig:RGImixing}. In addition to the works already described, this also
includes HPQCD09~\cite{Gamiz:2009ku} and
RBC/UKQCD14~\cite{Aoki:2014nga}. Results for the remaining 4 bag parameters in
the $\overline{\text{MS}}$ scheme at $\mu = m_b$ are shown in
Figure~\ref{fig:bags}.\footnote{The results from ETM for
  $\mathcal{B}_{B_q}^{(4)}$ and $\mathcal{B}_{B_q}^{(5)}$ have been recast into
  the same form as the results presented by HPQCD and FNAL/MILC, to guarantee
  that they are unity in the VSA. The required input for $M_{B_s}$, $M_{B_d}$,
  $\overline{m}_b(m_b)$, $\overline{m}_d(m_b)$ and $\overline{m}_s(m_b)$ were
  taken from Ref.~\cite{FermilabLattice:2016ipl}.}

The red left facing triangles representing the ETM $N_f=2$ result are shown with
open symbols, since the calculation does not quantify any uncertainty related to
the missing sea strange quark, which in many observables has been shown to be
significant. Since missing sea-charm effects are assumed to be negligible at
this level of precision, $N_f=2+1$ and $N_f=2+1+1$ results can be directly
compared and the magenta bands correspond to weighted averages of these. In
cases where $\chi^2/\mathrm{dof} \geq 1.25$, the dashed magenta lines indicate
the uncertainties obtained after the application of the PDG scale factor
$\sqrt{\chi^2/\mathrm{dof}}$. The results from FNAL/MILC and HPQCD dominate the
average, but show some tension. The $\chi^2/\mathrm{dof}$ for the weighted
average of just these two results is 1.2 and 2.4 for
$\mathcal{\hat{B}}^{(1)}_{B_d}$ and
$f_{B_d}(\mathcal{\hat{B}}^{(1)}_{B_d})^{1/2}$, respectively and larger than 3
for both quantities in the $B_s$ case.
It is noteworthy, that these tensions disappear when considering SU(3)-breaking
ratios, since the effect is correlated between the $B_d$ and the $B_s$ system.
The weighted averages of the observables shown in the right hand column of
Fig.~\ref{fig:RGImixing} have a precision of 2.1\%, 2.0\% and 0.8\%,
respectively. We note however, that if the average was only taken between the
FNAL/MILC and the HPQCD result, the uncertainties for the first two quantities
would be 3.5\% and 3.3\% due to the required rescaling of the uncertainties. The
uncertainty on the CKM matrix elements $\abs{V_{td}}$ and $\abs{V_{ts}}$ and
their ratio , which can be extracted by combining these observables with the
experimental measurements of $\Delta m_d$ and $\Delta m_s$, is clearly limited
by the theoretical uncertainty, due to the per mille-level precision of the
experimental inputs. This necessitates further theory improvements.

Turning our attention to the remaining bag parameters (see
Figure~\ref{fig:bags}) there is agreement between most results, but for
$\mathcal{B}^{(4)}_{B_q}$ and $\mathcal{B}^{(5)}_{B_q}$ we notice a slight
tension between the ETM and the other two results, whilst for
$\mathcal{B}^{(3)}_{B_q}$ we notice a tension between the FNAL/MILC and the
HPQCD results. The former is very similar to what is observed in neutral kaon
mixing~\cite{Garron:2016mva} and might be related to the choice of RI/MOM
instead of RI/SMOM.  Clearly further independent calculations are desirable to
address and resolve these tensions.

A joined effort between the RBC/UKQCD and the JLQCD
collaboration~\cite{Boyle:2021kqn,Tsang:2023lattice} aims to extend
Ref.~\cite{Boyle:2018knm} to predictions of the full set of operators in a fully
relativistic set-up. This is achieved by complementing the existing data set
described in the last column of Table~\ref{tab:mixing} by JLQCD domain wall
fermion ensembles, albeit with slightly different action parameters, with three
lattice spacings in the range $a\in [0.04,0.08]\,\mathrm{fm}$. This allows
simulations up to close to the physical $b$-quark mass, controlling the
remaining extrapolation to its physical value. The chiral extrapolation is
controlled by the inclusion of the physical pion mass ensembles of the RBC/UKQCD
collaboration. The use of chirally symmetric domain wall fermions throughout,
guarantees a fully non-perturbative renormalisation procedure with
continuum-like mixing pattern, addressing one of the main systematic
uncertainties present in current calculations.

\section{Recent developments \label{sec:recent}}
Many (actually all) of the PS to V transitions described in the Section on
semileptonic form factors are decays to multiple (two) hadrons in the final
state. The $D^*$ for example decays strongly into $D\pi$. Such effects are
typically included in the form factor chiral extrapolations through $\chi$PT or
HM$\chi$PT~\cite{Randall:1993qg}.  A formalism to treat such processes on the
lattice however has been introduced several years ago by Lellouch and L\"uscher
in Ref.~\cite{Lellouch:2000pv} and extended for the cases discussed here in
Ref.~\cite{Briceno:2014uqa}. First numerical
applications~\cite{Leskovec:2022ubd} have appeared only recently for the
$B\to\rho(\pi\pi) \ell \overline{\nu}$ process, which provides an alternative
exclusive channel for the extraction of $V_{ub}$. Further results, also for
other channels, can reasonably be expected in the near future.

The formalism relies on the relation between two-particle energy levels in a
finite volume and the infinite-volume scattering lengths~\cite{Luscher:1990ux}.
Since energy levels are directly continued to Euclidean time this allows to
compute properties of scattering processes on the lattice.  In a second step a
perturbation is introduced inducing the transition between a single particle and
two particle states ($K\to \pi\pi$ is the process considered in
Ref.~\cite{Lellouch:2000pv}). By computing (to leading order) the effect of such
a perturbation on the energy levels and the scattering, one obtains a
proportionality relation between the finite volume matrix element and the (say,
$K\to \pi\pi$) decay amplitude in infinite volume. The proportionality factor,
known as "Lellouch-L\"uscher" factor is a function of the momenta of the
particles and the derivative of the scattering phases with respect to them. In
Ref.~\cite{Briceno:2014uqa} a further step is taken by extending the approach to
the case of currents (the perturbations above) inserting energy, momentum and
angular momentum for systems with an arbitrary number of mixed two particle
states. In this case the application in mind is the $B\to K^*(K\pi) \ell^+
\ell^-$ transition.

Another topic in which considerable progress has been made, is the study of
inclusive decays on the lattice, with the aim to shed some light on the tensions
between inclusive and exclusive determinations of CKM matrix elements such as
$V_{cb}$ and $V_{ub}$.  In a series of papers~\cite{Hashimoto:2017wqo,
  Gambino:2020crt, Gambino:2022dvu,Barone:2023tbl} an approach has been devised, taking the
process $B_{(s)} \to X_c \ell \nu$ as prototype.  The central quantity is the
hadronic tensor
\begin{equation}
  \begin{aligned}
    W_{\mu\nu}(p_B,q)&=\sum_X(2\pi)^3 \delta(p_B-q-p_X) \times \\
    &\langle B(p_B)|J_\mu^\dagger|X\rangle\langle X| J_\nu |B(p_B)\rangle \;,
  \end{aligned}
\end{equation}
where $J_\mu$ is the weak current inducing the $b\to c $ transition, $q$ is the
lepton-pair momentum and the sum over the charmed final state $X$ includes a
spatial-momentum integral (in $\vec{p}_X$).  At fixed $p_B$, for example by
choosing the rest-frame of the $B$ meson, the tensor above is a function of the
spatial components $\vec{q}$ and $\omega=M_B-q_0$. What can be computed on the
lattice is essentially the Laplace Transform of such tensor at fixed $\vec{q}$
\begin{equation}
C_{\mu\nu}(\vec{q},t)=\int_0^\infty d \omega W_{\mu\nu}(\vec{q},\omega) e^{-\omega t}\;.
\end{equation}
Since $C_{\mu\nu}$ in finite volume is given by a sum of exponentially falling
(in time) functions, trying to invert the relation above for arbitrary values of
$\omega$ is an ill-posed problem. For $\omega$ smaller than the energy of the
lowest state $X$ the relation however can be inverted. In
Ref.~\cite{Hashimoto:2017wqo} the tensor $W_{\mu\nu}(\vec{q},\omega)$ is
computed on the lattice for that particular unphysical kinematic choice ($\omega
< E_X^\mathrm{min}$), where the final hadronic state can not go on shell and is
then related to derivatives of the tensor in the physical region through a
dispersion integral. The approach is very similar to the derivation of the
moment sum rules in the case of Deep Inelastic Scattering.

Improving on this first approach a completely new method has been introduced in
Ref.~\cite{Gambino:2020crt}, based on the observation that in order to compute
the inclusive decay rate what is needed is not the hadronic tensor but rather a
smeared version of it with functions resulting from the leptonic tensor. The
building blocks are the quantities
\begin{equation}
\bar{X}^{(l)}(|\vec{q}|^2)=\int_0^\infty d \omega W_{\mu\nu}(\vec{q},\omega)K^{\mu\nu,(l)}(\vec{q},\omega) \;,
\end{equation}
where the functions $K^{\mu\nu,(l)}(\vec{q},\omega)$ are known and defined in
Ref.~\cite{Gambino:2020crt}. If one can find good polynomial approximations of
the $K$-functions in powers of $e^{-a\omega}$, then the $\bar{X}^{(l)}$ can be
obtained by taking suitable combinations of the correlator $C_{\mu\nu}$ at
different times. Obviously the larger the time one can compute the correlation
function with controlled errors, the higher the degree one can consider the
polynomial approximation and the better the accuracy in
$\bar{X}^{(l)}(|\vec{q}|^2)$, and eventually in the inclusive decay rate.  In
Ref.~\cite{Gambino:2020crt} Chebyshev polynomials have been studied while in
Refs.~\cite{Gambino:2022dvu,Barone:2023tbl}, in an attempt to optimise the approximation
balancing truncation errors and statistical noise from the correlator
$C_{\mu\nu}$, also the Backus-Gilbert method has been considered.  All such
studies are still exploratory, however the results are very encouraging.

\section{Conclusions and outlook\label{sec:conc}}
The combination of Lattice QCD and $B$-physics constitutes an exciting and
active field. In this review, we have commented on the challenges faced when
making predictions for $B$-physics observables from lattice QCD calculation. We
have reviewed the recent literature for the determination of the $b$-quark mass,
leptonic decays constants, several semileptonic decay channels and neutral
meson mixing parameters. To address some of the observed tensions, we have put
forward a number of suggestions for benchmark quantities, which will help to
understand systematic effects associated with complicated multi-dimensional
fits.

Some quantities ($b$-quark mass, decay constants) are in a mature state,
with many results using different methodologies agreeing, producing comparable
uncertainties, and the corresponding tests of the SM are limited by the
experimental precision. The more complicated calculation of semileptonic form
factors has made tremendous progress, but due to to the wealth of decay channels
and the increased computational and data-analysis complexity, fewer independent
results exist for the individual observables. Several of the results in the
literature have not yet reached a community consensus and currently display some
level of tension. However, there are major ongoing efforts by several groups to
consolidate these calculations. We hope that the benchmark quantities we suggest
in Sec.~\ref{subsec:benchmarks} will be adopted by the community and will aid in
shedding light on current tensions between different lattice computations. The
situation for neutral meson mixing parameters is similar with only a small
number of results currently available.

Finally, we comment on recent progress in calculations of semileptonic decays into
two final state hadrons and inclusive semileptonic decays. Whilst these
calculations are currently exploratory they are very promising and we expect
they will mature to further ab-initio tests of the SM.

It is worth emphasising the increase in complexity in the quantities presented
here.  Masses and decay constants can be extracted from two-point functions,
form factors require three-point functions while for two-hadron decays and
inclusive transition rates four-point functions are needed. The statistical
noise in $n$-point functions unavoidably grows with $n$ and in addition the
combinatorics in the number of Wick contractions becomes more complex. It is a
remarkable achievement of the advances in numerical methods and computer power
combined with smart field theoretical ideas that has led us to tackle new
challenges in $B$-physics on the lattice, which seemed un-treatable only a few
years ago.
\section*{Acknowledgements}
We thank Daniel Wyler for inviting us to contribute to the ``European Physical
Journal Special Topics on B-physics'' volume. We are grateful to Marzia Bordone
and Shoji Hashimoto for comments on the manuscript.

\vspace{0.3cm}
\noindent Data Availability Statement: No Data associated in the manuscript.
{\small
  \bibliographystyle{JHEP-notitle}
  \bibliography{review.bib}
}

\end{document}